\documentclass[aps,prd,preprint,nofootinbib,longbibliography,superscriptaddress]{revtex4-1}

\usepackage{amsmath}
\usepackage{amssymb} 
\usepackage[cm]{fullpage}
\usepackage{graphicx} 
\usepackage{siunitx}
\usepackage{tabularx} 
\usepackage[colorlinks,linkcolor=blue,citecolor=blue,urlcolor=blue]{hyperref}
\usepackage[table,xcdraw]{xcolor}
\usepackage[utf8]{inputenc}
\usepackage{multirow}

\DeclareSIUnit\parsec{pc}

\newcommand{\beq}{\begin{equation}}
\newcommand{\eeq}{\end{equation}}

\newcommand{\rpar}[1]{\left(#1\right)}
\newcommand{\spar}[1]{\left[#1\right]}

\newcommand{\bd}{{\mathrm d}}
\newcommand{\mpl}{M_\text{Pl}}

\newcommand{\ip}{\text{ip}}
\newcommand{\eq}{\text{gb}}
\newcommand{\de}{\text{de}}

\newcommand{\rhophib}{\rho_{\phi,\text{b}}}
\newcommand{\rhophia}{\rho_{\phi,\text{a}}}

\newcommand{\m}{q}
\newcommand{\n}{p}

\begin{document}

\title{Gauss-Bonnet-coupled Quintessential Inflation}
\date{\today}
\author{Carsten van de Bruck}
\affiliation{Consortium for Fundamental Physics, School of Mathematics and Statistics, University of Sheffield, Hounsfield Road, Sheffield S3 7RH, United Kingdom}
\author{Konstantinos Dimopoulos}
\affiliation{Consortium for Fundamental Physics, Physics Department,Lancaster University, Lancaster LA1 4YB, United Kingdom}
\author{Chris Longden}
\affiliation{Consortium for Fundamental Physics, School of Mathematics and Statistics, University of Sheffield, Hounsfield Road, Sheffield S3 7RH, United Kingdom}
\author{Charlotte Owen}
\affiliation{Consortium for Fundamental Physics, Physics Department,Lancaster University, Lancaster LA1 4YB, United Kingdom}

\begin{abstract}
We study in detail a new model of quintessential inflation where the inflaton 
field is coupled to the Gauss-Bonnet term. This coupling ensures that the 
variation of the field is kept sub-Planckian, which avoids the 5th force problem
as well as the lifting of the flatness of the quintessential tail in the runaway scalar
potential due to radiative corrections. We find that the inflationary 
predictions of the model are in excellent agreement with CMB observations, 
while the coincidence requirement of dark energy is satisfied with natural 
values of the parameters, overcoming thereby the extreme fine-tuning of the 
cosmological constant in $\Lambda$CDM.
\end{abstract}

\maketitle

\tableofcontents

\pagebreak

\section{Introduction} 

Cosmic inflation is widely accepted as the leading paradigm for describing the 
physics of the very early Universe. It entails an accelerating expansion of 
space in the earliest moments of the Universe's history, alleviating the 
infamous horizon and flatness problems of the Hot Big Bang, while also 
generating the near scale-invariant spectrum of initial curvature perturbations 
necessary to seed the anisotropies in  the Cosmic Microwave Background (CMB) as 
probed by observational tests such as WMAP~\cite{Bennett:2012zja} and 
Planck~\cite{Ade:2015tva,Planck2015Inflation}. Another problem in cosmology, 
however, is the observed present accelerated expansion of space, as identified 
by type Ia supernova distance measurements \cite{Perlmutter:1998np} as well as CMB physics. From these, we can deduce that the Universe's present accelerated expansion is consistent with the effects of a perfect fluid (dark energy) with equation of 
state parameter close to $-1$, that comprises some 70~\% of the Universe's  
density~\cite{Ade:2015xua}. While many approaches to solving these two problems 
in cosmology are largely independent of one another, it seems inevitable that 
cosmologists would try to unify explanations of these similar periods of 
accelerated expansion in one theory. Such an approach is what is known as 
quintessential inflation \cite{Peebles:1998qn,Dimopoulos:2000md,Dimopoulos:2001ix,Ng:2001hs,Huey:2001ae,Majumdar:2001mm,Nunes:2002wz,Rosenfeld:2005mt,Rosenfeld:2006hs,BasteroGil:2009eb,Bento:2008yx,Cardenas:2006py,Dias:2010rg,Frewin:1993aj,Geng:2015fla,Geng:2017mic,Guendelman:2016kwj,deHaro:2016ftq,Haro:2015ljc,deHaro:2017nui,Hossain:2014xha,Hossain:2014zma,Khalifeh:2015tla,Membiela:2006rj,Neupane:2007mu,Rubio:2017gty,Sami:2004xk,BuenoSanchez:2006fhh,BuenoSanchez:2006epu,Zhai:2005ub,Ahmad:2017itq,Peebles:1998qn,Bento2009}.

One of the main similarities between many approaches to understanding inflation 
and dark energy is the exploitation of a scalar field, $\phi$, which has 
properties conducive to accelerating expansions of space. In theories which 
deal only with inflation, this \emph{inflaton} field typically decays into the 
standard model particles once it has served its purpose. This has the advantage 
of simply recovering the conventional radiation and matter dominated epochs of 
the Hot Big Bang model, a so-called graceful exit from inflation, but leaves us 
with no choice but to later invoke a second new mechanism to generate dark 
energy. If instead, the inflaton at least partly persists following the
end of inflation, it is feasible that it may eventually come to dominate the 
energy budget of the Universe again and give rise to dark energy. 

While this idea sounds simple in principle, achieving it is theoretically 
challenging. For one thing, the energy scale of inflation and that of dark 
energy differ by more than a hundred orders of magnitude. Manually introducing 
an energy scale as small as that of dark energy into one's theory renders it 
hopelessly unstable under quantum corrections, and if one instead chooses to 
suppress the energy density of the inflaton by having it roll down a steep 
potential following inflation, one finds that a trans-Planckian excursion in 
field value typically occurs before the field freezes in at a particular value, 
spoiling the stability of the theory under UV completion. This latter approach 
also makes recovering the standard cosmology slightly more complicated because
one needs to reheat the Universe after inflation by means other than the decay 
of the inflaton field. However, this is not intractable thanks to decay 
procedures such as instant preheating \cite{Felder1998,Felder:1999pv,Campos:2002yk},
curvaton reheating \cite{Feng:2002nb,Matsuda:2007ax,BuenoSanchez:2007jxm}
or gravitational reheating \cite{Chun:2009yu,Ford:1986sy}
, which we will also apply in this work.

In this paper we seek to study the feasibility of quintessential inflation 
scenarios in a modified theory of gravity in which the scalar field in the 
theory non-minimally couples to the Gauss-Bonnet combination of quadratic 
curvature scalars, $R^2-4R^{\mu\nu}R_{\mu\nu}+R^{\rho\mu\sigma\nu}R_{\rho\mu\sigma\nu}$. 
Scalar-tensor theories with such a Gauss-Bonnet coupling (and to a lesser 
extent, vector-tensor theories~\cite{Oliveros:2016myr}) have been studied 
extensively with applications to inflation~\cite{Tsujikawa:2004dm,Guo:2010jr,Jiang:2013gza,Koh:2014bka,vandeBruck:2015gjd,vandeBruck:2016xvt,Koh:2016abf,Mathew:2016anx,Okada:2014eva,Fomin:2017qta} and dark energy~\cite{Linder:2017egp, Santillan:2017nik} investigated separately, as well as topics such as black hole 
formation~\cite{Sotiriou:2013qea} . This is unsurprising, as it is a 
well-motivated extension of General Relativity, appearing in UV theories such as
string~\cite{Kawai:1998ab,Kawai:1999pw} and braneworld-inspired~\cite{Binetruy:2002ck,Brax:2003fv,Charmousis:2002rc,Germani:2002pt,Gravanis:2002wy} models, as 
well as just being a fairly natural object to consider when building gravity 
theories from the bottom up~\cite{Donoghue:2012zc}, as the simplest curvature 
scalar which does not add any additional propagating degrees of freedom to the 
theory as, say, $R^2$ or $R_{\mu \nu} R^{\mu \nu}$ alone do. Realisations of 
bouncing cosmologies have also been found in Gauss-Bonnet-containing 
theories~\cite{Bamba:2014mya,Gasperini:1996fu}. It is also a subset of 
Horndeski's theory~\cite{Horndeski:1974wa,Charmousis:2011ea,Copeland:2012qf}, 
guaranteeing that it has second order equations of motion and is free of 
instabilities.

A particular reason we are interested in the application of Gauss-Bonnet-coupled
theories to quintessential inflation is because previous work~\cite{vandeBruck:2016xvt,vandeBruck:2015gjd,Kawai:1998ab} on fields with such a coupling has 
revealed that a common behaviour resulting from this is the impedance of motion 
of the coupled scalar field when the Gauss-Bonnet coupling function becomes 
large. This may be especially useful in overcoming the aforementioned 
theoretical problem where a field rolling down a quintessential potential tail 
after inflation becomes super-Planckian in displacement.

In section~\ref{sec:model} we will begin by specifying the concrete model we 
will use to study these questions about the realisation of quintessential 
inflation in Gauss-Bonnet-coupled models. We will discuss the slow-roll 
inflationary dynamics and power spectra predictions in 
section~\ref{sec:model_inflation}, before moving on to considerations of how 
reheating is subsequently achieved in section~\ref{sec:model_reheating}, 
retrieving a Hot Big Bang universe. Then, the late-time behaviour and 
manifestation of dark energy from the leftover inflaton density will be 
described in section~\ref{sec:model_DE}. Having shown how our model allows one 
to realise these three key steps in quintessential inflation, we will proceed in
section~\ref{sec:constraints} to use our results to assess the quantitative 
feasibility of the model by scrutinising its predictions against the 
observational evidence, deriving any resulting constraints on the parameter 
space of the model. In particular, this will involve constraints from inflation 
in section~\ref{sec:constraints_inflation}, reheating and dark energy in 
section~\ref{sec:constraints_reheating_DE}, and lastly, local limits on modified
gravity models in section~\ref{sec:constraints_tests}. We then finish with some 
concluding remarks in section~\ref{sec:conc}. 

We consider natural units, where $c=\hbar=1$ and Newton's gravitational 
constant is \mbox{$8\pi G=M_{\rm Pl}^{-2}$}, with 
\mbox{$M_{\rm Pl}=2.43\times 10^{18}\,$GeV} being the reduced Planck mass.

\section{The Model} \label{sec:model}

Consider a theory of modified gravity in which a scalar field, $\phi$, is 
non-minimally coupled to the Gauss-Bonnet combination of quadratic curvature 
scalars, $E=R^2-4R^{\mu\nu} R_{\mu \nu} + R^{\rho \mu \sigma \nu} R_{\rho \mu \sigma \nu}$, 
such that
\beq \label{eq:GBaction}
S = \frac{\mpl^2}{2} \int \bd^4 x \sqrt{-g} \spar{ R - G(\phi) E } 
-\int \bd^4 x \sqrt{-g} \spar{\frac{1}{2} (\partial \phi)^2 + V(\phi)  } \, .
\eeq
While the Gauss-Bonnet combination is usually a total derivative that has no 
effect on the classical equations of motion, the coupling with the scalar field 
$G(\phi)$ in the above action, as long as it is a non-constant function, will 
allow it to play a non-trivial role. Considering the metric of a spatially flat,
homogeneous and isotropic spacetime with 
\begin{align}
ds^2 = -dt^2 + a^2(t)\delta_{ij}dx^i dx^j, 
\end{align}
where $a(t)$ is the scale factor,
the equations of motion derived from this action are
\begin{align}
3 \mpl^2 H^2 & = \frac{1}{2} \dot{\phi}^2 + V(\phi) +12 \mpl^2 H^3 \dot{G} \, , 
\label{eq:FriedmanGB1} \\
2 \mpl^2 \dot{H} & = - \dot{\phi}^2 + 4 \mpl^2  H^2 (\ddot{G}-H\dot{G}) + 
8 \mpl^2  H \dot{H} \dot{G} \,  ,  \label{eq:FriedmanGB2} \\
\ddot{\phi} & + 3 H \dot{\phi} + V_{,\phi} + 12 \mpl^2  H^2 G_{,\phi} 
(\dot{H} + H^2) = 0  \, ,  \label{eq:GBKGE}
\end{align}
where the dot denotes a time derivative.

To achieve quintessential inflation in this model, we want the new Gauss-Bonnet 
effects to play a significant role at late times, while allowing for an 
inflationary expansion (e.g. a sufficiently flat effective potential) at early 
times. To this effect, we choose a fairly minimalistic coupling function of the 
form
\beq
G(\phi) = G_0 e^{-\m \phi / \mpl} \, , \quad \m > 0 \, ,
\label{GB}
\eeq
where the prefactor should satisfy $G_0 \mpl^2 \geq 1$ in a sub-Planckian 
theory. For large positive $\m\phi$ values, this will be typically negligible, 
while for negative $\m \phi$ it will quickly grow in magnitude. Thus, in a 
scenario in which $\m \phi$ begins at large positive values and rolls down to 
negative values at late times, this coupling may behave as needed. The other 
component of the model, the potential, is consequently taken to be
\beq
V(\phi) = \frac{V_0}{2} \spar{1 + \tanh \rpar{\n \;
\frac{\phi - \phi_c}{\mpl}}} \, , \quad \n > 0 \, . \label{eq:potentialtanh}
\eeq
We have chosen this potential as a mathematically convenient prototype for 
situations in which an early time plateau, favoured by Planck, as well as an 
exponential quintessential tail are present, while remaining agnostic as to its 
origin. We expect our results to qualitatively hold even if the precise form of 
the potential is changed, so long as at late times there is a plateau suitable 
for quintessential inflation and that at early times inflation may be realised. 
Note also, that after inflation, the field becomes kinetically dominated and 
oblivious of the potential until it eventually freezes somewhere along the quintessential 
tail.

One could also consider scenarios in which the Gauss-Bonnet coupling is not 
negligible at early times, and the inflation-determining effective potential is 
due to a mixture of $V$ and $G$ \cite{Guo:2010jr}, but for simplicity, we will not consider this 
in detail. Note that the value of the constant $\phi_c$, under a field 
redefinition $\phi \rightarrow \phi + \phi_c$, can be absorbed into a rescaling 
of the constant $G_0 \rightarrow G_0 e^{\m \phi_c}$, and so we can (and henceforth
will) set it to zero without loss of generality.

\subsection{Inflation} \label{sec:model_inflation}
As the coupling to the Gauss-Bonnet term is assumed unimportant during inflation
($\m \phi \gg 1$), unlike in e.g.  \cite{Guo:2010jr} where both the potential 
and the GB coupling play a role in the inflationary dynamics, we can proceed to 
apply the usual slow-roll formalism to study inflation. We find that the 
slow-roll parameter $\epsilon = - \dot{H}/H^2$ is 

\beq
\epsilon \simeq \frac{\mpl^2}{2} \rpar{\frac{V_{\phi}}{V}}^2 = 
\frac{\n^2}{2} \spar{1 - \tanh \rpar{\n\;
\frac{\phi}{\mpl}}}^2 \, ,
\eeq
where $\simeq$ denotes the slow-roll approximation. Inflation ends when 
$\epsilon = 1$, which results in
\beq
\phi_\text{end} \simeq \frac{\mpl}{\n} \tanh^{-1} \rpar{1 - \frac{\sqrt{2}}{\n}} \, .
\eeq
The $e$-folding number is then found to be

\begin{align}
N & = \frac{1}{\mpl} \int^{\phi}_{\phi_\text{end}} 
\frac{ \bd \phi }{\sqrt{2 \epsilon}} \, \nonumber \\ & 
\simeq \frac{1}{4\n^2} e^{2 \n \phi/\mpl} + \frac{\phi}{2\n \mpl} - 
\frac{1}{2\n^2} \spar{\frac{\n}{\sqrt{2}} + 
\tanh^{-1} \rpar{1 - \frac{\sqrt{2}}{\n}} - \frac{1}{2}}\, , 
\label{eq:efoldssra}
\end{align}
The first term here is dominant, and so inverting Eq.~(\ref{eq:efoldssra}) we 
obtain the approximate initial condition for $N$ $e$-folds of inflation to subsequently occur as

\beq
\phi(N) \approx \frac{\mpl}{2\n} \log 4\n^2 N \, .
\eeq
We can obtain an expression for the power spectrum in terms of $N_*$, the number of e-folds remaining until the end of inflation when the cosmological scales exit the horizon

\beq
\mathcal{P}_\mathcal{R} \simeq \frac{H^2}{8 \pi^2 \epsilon} \simeq  
\frac{\n^2 V_0 N_*^2}{3 \pi^2 \mpl^4} \, , \label{i_ppsa}
\eeq
where the 
value of $N_*$ is typically about $60$ for observable scales. We proceed further
to find the spectral index
\beq
n_s - 1 \simeq 
- \frac{4\n^2 \rpar{1+8 \n^2 N_*}}{\rpar{1 + 4 \n^2 N_*}^2} \approx 
-\frac{2}{N_*} \, , \label{eq:i_ns}
\eeq
and the tensor-to-scalar-ratio
\beq
r \simeq 16 \epsilon \simeq 
\frac{32 \n^2}{\rpar{1 + 4\n^2 N_*}^2} \approx 
\frac{2}{\n^2 N_*^2} \, . \label{eq:i_ttsr}
\eeq

\subsection{Reheating}  \label{sec:model_reheating}
First, consider the post-inflationary evolution of the system, ignoring 
radiation. A static ($\phi = $ constant) solution of the equations of motion 
(Eqs. (\ref{eq:FriedmanGB1}) -- (\ref{eq:GBKGE})) with 
$\dot{\phi} = \ddot{\phi} = \dot{H} = 0$ exists, as one expects from previous 
work indicating that one function of a large Gauss-Bonnet coupling is freezing 
the time evolution of the inflaton \cite{vandeBruck:2015gjd}. One finds that 
the constant value the field approaches is given by

\beq
\phi_s / \mpl \approx  \frac{1}{\m - 2\n}\ln \alpha \, , 
\label{eq:static_solution}
\eeq
where
\beq
\alpha \equiv \frac{2 \m V_0 G_0}{3\n \mpl^2} \, .
\eeq
Numerically, we observe that this solution is approached in the 
post-inflationary regime as the Gauss-Bonnet term becomes important, at negative
field values. As the field is frozen, this solution itself is an inflationary 
expansion with $\epsilon = 0$. This implies an unsuitability for perturbative 
reheating \cite{Kofman:1997yn}, as found in previous GB-coupled models 
\cite{vandeBruck:2016xvt}, as no oscillatory behaviour about a potential 
minimum exists. Between the initial period of inflation, and this late-time 
accelerating expansion, however, there is an interval when the field is rolling 
quickly down the steep decline of the potential around $\phi = 0$. Around this 
point, it is feasible to implement instant preheating 
\cite{Felder1998,Felder:1999pv,Campos:2002yk}

to recover a radiation-dominated epoch. For concreteness, we consider a coupling
between the inflaton and a matter field $\chi$ (which is assumed to subsequently
decay efficiently into radiation) of the form
\beq \label{eq:RehLagrangian}
\mathcal{L} = -\frac{1}{2} m_{\chi,0}^2 \chi^2 - 
\frac{1}{2} g^2 (\phi - \nu)^2 \chi^2 .  
\eeq
It is well known that this approach, if one takes the bare mass $m_{\chi,0}$ to 
be small compared to the induced mass $g |\phi-\nu|$,  leads to the production 
of $\chi$ particles with total energy density
\beq \label{eq:IPlimit1}
\rho_\chi = 
\frac{g^{5/2}|\dot{\phi}|_{\phi_\ip}^{3/2} (\nu - \phi_\ip)}{8 \pi^3} \, ,
\eeq
where $\phi_\text{ip}$ is the $\phi$ value at the time $t_\ip$ of instant 
preheating ($\phi_\text{ip}\equiv\phi(t_\ip)$). Furthermore, we choose $\nu = 0$ in the spirit of minimalism, though instead choosing a small non-zero value is not expected to significantly affect our results. In instant preheating, particle 
production occurs explosively around the time when the non-adiabaticity 
condition, $|\dot{m}_\chi| > m_\chi^2$, where $m_\chi \approx g |\phi-\nu|$, is 
first satisfied. We hence take this to be the time of instant preheating for 
this purpose, and can determine when this occurs via a numerical integration of 
the equations of motion (\ref{eq:FriedmanGB1})--(\ref{eq:GBKGE}) for a short 
time after the end of inflation.

For instant preheating to induce radiation domination, it is necessary that 
$\rho_\chi$ is greater than $\rho_\phi$ after instant preheating. Denoting the 
energy density of $\phi$ before and after instant preheating occurs as 
$\rhophib$ and $\rhophia$, respectively, we hence impose

\beq
\rho_\chi > \rhophia \quad 
\Rightarrow
\quad \rho_\chi > \frac{1}{2} \rhophib \, ,
\eeq
where by energy conservation we require $\rhophia = \rhophib - \rho_\chi$.

After instant preheating, we also want the $\phi$ field's dynamics to be 
dominated by its kinetic energy density, because a potential-dominated inflaton 
field will quickly come to dominate again, thereby terminating the 
radiation-dominated epoch. As a result, we wish for the kinetic energy density 
of the inflaton after instant preheating to be greater than its potential. 
Considering that the potential remains constant\footnote{%
Note that during instant preheating, it is purely kinetic energy density that is
converted to radiation.}
throughout instant preheating (i.e. $V(\phi_\ip) = V_\text{a} = V_\text{b}$), this
means we want
 
\beq \label{eq:IPlimit2}
\rhophia - V(\phi_\ip) > V(\phi_\ip) \quad 
\Rightarrow
\quad \rho_\chi < \rhophib - 2 V(\phi_\ip) \, ,
\eeq
where again we considered energy conservation. Combining the inequalities in 
Eqs.~(\ref{eq:IPlimit2}) and (\ref{eq:IPlimit1}), we hence obtain the range of 
suitable $\rho_\chi$ values as

\beq \label{eq:reheatingconditions}
\frac{1}{2} \rhophib < \rho_\chi < \rhophib - 2 V(\phi_\ip) \, .
\eeq
This result implies that the implementation of instant preheating will only be 
able to succeed when it occurs at a sufficiently kinetic-dominated moment in the
evolution of the inflaton.
From Eq.~\eqref{eq:reheatingconditions}, we find the constraint

\beq \label{eq:kinetic_potential_constraint}
\rho_{\mathrm{kin,b}} > 3 V(\phi_\ip)\,.
\eeq
The potential must hence be sufficiently steep that the field rolls fairly 
quickly after inflation. This will be discussed in more detail in 
section~\ref{sec:constraints_reheating_DE}.

\subsection{Behaviour in Radiation/Matter Dominated Epochs} 
\label{sec:model_DE} 

After reheating has occurred according to the details of 
section~\ref{sec:model_reheating}, the Universe becomes radiation-, or eventually
matter-, dominated. In such regimes, we have $H = k /t$, where $k$ is a constant 
depending on the epoch in question (in particular, $k=1/2$ for a 
radiation-dominated Universe and $k=2/3$ for a matter-dominated Universe). We 
assume that immediately after instant preheating, the Gauss-Bonnet term is still
negligible. This is desirable because if the field were to become GB-dominated, 
and hence freeze, immediately after reheating, its density would be too large to
act as dark energy. Instead, we require that that GB coupling only becomes 
significant at some, as of yet unspecified, later time, when the field has 
rolled further down the quintessential tail to reduce its final energy density. 
To achieve this in a sub-Planckian field displacement, we expect to need a 
fairly large $\n$ value to make the quintessential tail of the potential steep 
enough to suppress the energy density by many orders of magnitude during this 
process. Ordinarily such a steep tail would not lead to accelerated expansion, 
hence the necessity of the Gauss-Bonnet term.

Due to the requirements set out in Eq.~(\ref{eq:reheatingconditions}), the field
will necessarily be kinetically dominated immediately after instant preheating. 
We can hence neglect the potential and GB terms in the equation of motion, and 
determine the subsequent evolution of the field via

\beq \label{eq:kinatiionKGE}
\ddot{\phi} + 3 H \dot{\phi} \simeq 0 \, .
\eeq
In conventional quintessential inflation \cite{Dimopoulos:2001ix,Dimopoulos:2017zvq}, 
the solution of this equation determines the late time behaviour of the field, 
and it is known that the field eventually freezes in at a certain value, but 
only after undergoing a super-Planckian displacement. This freezing behaviour 
is conducive to achieving late-time dark energy, but, as the field has become 
super-Planckian, radiative corrections mean the flatness of the potential 
cannot be guaranteed, the field may act as a `5th force' violating the 
equivalence principle and the model is too sensitive to the details of UV 
completion to be trusted \cite{Wetterich2004,Carroll:1998zi}.

In our model, however, as the field rolls to more negative values, the size of 
the Gauss-Bonnet coupling will grow exponentially and eventually become 
non-negligible. Eq.~(\ref{eq:kinatiionKGE}) is hence only valid up until a time 
when the contribution of the GB coupling to the Klein-Gordon equation becomes 
comparable to $\ddot{\phi}$. To solve the full nonlinear ODE of 
Eq.~(\ref{eq:GBKGE}), even when neglecting the potential, is difficult if not 
impossible, so here we instead resort to an approximation where immediately 
following instant preheating, the GB coupling is still negligible and 
Eq.~(\ref{eq:kinatiionKGE}) determines the evolution of the system (the 
kinetic-dominated regime). 

This holds until a time $t_\eq$, when $|\ddot{\phi}|=|12 \mpl^2 H^2(\dot{H}+H^2)G_{,\phi}|$. 
After this, the second derivative term in the Klein-Gordon equation is instead 
treated as negligible while the GB contribution is accounted for (the 
GB-dominated regime), and the system's evolution is given by solutions to

\beq \label{eq:GBdomKGE}
3 H \dot{\phi} + 12 \mpl^2 H^2 G_{,\phi} (\dot{H} + H^2) = 0  \, . 
\eeq

The two regimes' solutions will then be ``stitched'' together via the boundary 
condition that the two solutions must agree at $t_\eq$. With this method we lose 
the ability to finely resolve the evolution of $\phi$ about $t_\eq$, but should 
still be able to determine the all-important late time behaviour with reasonable accuracy.

\subsubsection{Kinetic Regime}

We fix the initial conditions to solve Eq.~(\ref{eq:kinatiionKGE}) as 
$\phi(t_\text{ip}) = \phi_\text{ip}$ and $\dot{\phi}(t_\text{ip}) = 
\dot{\phi}_\text{ip} = -\sqrt{6 \Omega_\text{ip}}( k / t_\text{ip}) \mpl$. The 
first initial condition merely imposes that the field begins at the value it had
when instant preheating occurred, while the second initial condition uses the 
first Friedman equation (\ref{eq:FriedmanGB1}) to set $\dot{\phi}$ at this time,
as a function of $\Omega_\ip \equiv \Omega_\phi(t_\ip)$ - the density parameter of
the field at the moment of instant preheating (immediately after). We also 
assume that the field rolls towards negative values. The discussion in 
section~\ref{sec:model_reheating}, where we require that the produced radiation 
density dominates over the remaining field density, implies that 
$\Omega_\ip\ll 1$. Solving Eq.~\eqref{eq:kinatiionKGE} with these conditions 
gives

\beq
\phi(t) = \phi_\text{ip} - \mpl \sqrt{6 \Omega_\text{ip}}\rpar{\frac{k}{3k - 1}}
\spar{1 - \rpar{\frac{t_\text{ip}}{t}}^{3k-1}} \, . \quad (t < t_\eq) 
\label{eq:raddomphi}
\eeq

\subsubsection{Gauss-Bonnet Regime}

As the GB-dominated equation of motion, Eq.\eqref{eq:GBdomKGE}, is first order, we 
only need the initial condition that at the time when the GB-dominated solution 
first becomes relevant $t_\eq$, the field takes the value (which will later be 
determined) $\phi_\eq = \phi(t_\eq)$. Using this, we find the solution

\beq
\phi(t) = \phi_\eq + \frac{\mpl}{\m} \ln \spar{1 + 2 G_0 \m^2 k^2 (1-k) 
e^{-\m \phi_\eq/\mpl} \rpar{\frac{1}{t^2}-\frac{1}{t_\text{\eq}^2}}} \, .  
\label{eq:gbdomphi}
\eeq
At very late times $(t \gg t_\eq)$ the field will then tend to a constant value
\beq
\phi(t \gg t_\eq) = \phi_\eq + \frac{\mpl}{\m} \ln \rpar{1 - \frac{\beta G_0 
\m^2 e^{-\m \phi_\eq/\mpl} }{t_\eq^2}} \approx \phi_\eq \, , \quad (t > t_\eq)  
\label{eq:gbdomphi2}
\eeq
where $\beta = 2 k^2 (1-k)$. Hence $\beta= 1/4$ in the radiation dominated case 
and $\beta = 8/27$ for the matter dominated case. In the second approximate 
equality, we note that for typical parameter values, there is very little 
variation of the field in this regime as the second term is generally quite 
small. This is expected, as the principle of our model is that a large GB 
coupling impedes the evolution of the field so $\phi$ freezes almost immediately
when GB becomes important.

\subsubsection{Stitching and Boundary Condition}\label{sec:stitch_sol}

Having determined the evolution of the field for $t < t_\eq$, 
Eq.~(\ref{eq:raddomphi}), and $t > t_\eq$, Eq.~(\ref{eq:gbdomphi}), we need to now
determine the moment $t_\eq$ at which these two solutions coalesce. As discussed
previously, this is when $|\ddot{\phi}| = |12 \mpl^2 H^2 (\dot{H}+H^2) G_{,\phi}|$. We
hence substitute the function $\phi(t)$ from Eq.~(\ref{eq:raddomphi}) into this 
condition and solve for the time at which it is first met. Doing so we obtain an
equation of the form

\beq
A t^\nu = \exp \rpar{-B t^\mu} \, ,
\eeq
which can be solved with the Lambert $W$ function, which satisfies 
$x = W(x) e^{W(x)}$, as
\beq
t_\eq = \spar{\frac{\nu}{B \mu} W\rpar{\frac{B \mu}{\nu} \rpar{\frac{1}{A}}^\frac{\mu}{\nu}}}^\frac{1}{\mu} \, , \label{eq:stitchingeqsol}
\eeq
where
\begin{align}
A & = \frac{1}{2 \m k G_0 (1 - k)} \sqrt{\frac{3 \Omega_\ip}{2}} \exp \rpar{\m \phi_\text{ip}/\mpl - \frac{\m k \sqrt{6 \Omega_\ip} }{3k - 1}} t_\text{ip}^{3k-1} \, , \\
B & = \m k \sqrt{6 \Omega_\ip}\rpar{\frac{t_\text{ip}^{3k-1}}{3k - 1}}\, , \\
\mu & = 1 - 3k \, , \\
\nu & = 3 - 3k \, .
\end{align}
In the region where the Lambert function $W(x)$ has two branches 
$(-e^{-1} < x < 0)$, this would imply there are two times at which the GB and 
second derivative contributions to the Klein-Gordon equation are equal, but of 
course only the earlier time of the two solutions is valid, provided that it 
obeys $t_\eq > t_\text{\ip}$, as Eq.~(\ref{eq:raddomphi}) is only valid up 
until the GB contribution first becomes important. Typically it is the lower 
($W_{-1}$) branch of the function which evaluates to the relevant value, but in 
cases such as those where the lower branch yields $t_\eq < t_\text{ip}$, as it is 
physically ruled out (instant preheating must happen before GB-domination else 
a viable late-time universe is not recovered), we instead use the principal 
($W_0$) branch solution.

There may be parameter space in which there is no real solution of this 
equation, corresponding physically to there being no time of equality between 
these two terms in the Klein-Gordon equation. This would either imply that the 
GB term is already dominant at $t_\ip$, or that the field remains 
kinetic-dominated forever. Both cases are undesirable as they do not correspond 
to late-time dark energy (either the field freezes due to GB too soon after 
inflation to reduce its density to that of dark energy, or conventional 
quintessential inflation is recovered as GB is irrelevant). The reality of 
Eq.~\eqref{eq:stitchingeqsol} is hence an important check that we are looking at
feasible models, in particular, the argument of the Lambert function must 
satisfy
\beq
\frac{B \mu}{\nu} \rpar{\frac{1}{A}}^\frac{\mu}{\nu} \geq -\frac{1}{e} \, ,
\eeq
to have at least one real value.

Substituting the value of $t_\eq$ into Eq.~(\ref{eq:raddomphi}) allows us to 
determine $\phi_\eq$, and in turn this allows us to determine $\phi(t)$ for 
$t \gg t_\eq$ by substituting that into Eq.~(\ref{eq:gbdomphi2}). Doing so, we 
obtain
\beq
\phi_m = \phi(t \gg t_\eq) \approx \phi_\text{ip} + \frac{\mpl B}{\m} 
\rpar{t_\eq^\mu - t_\text{ip}^\mu}  + \frac{\mpl}{\m} 
\ln \rpar{1+ \frac{\mu B}{2} t_\eq^\mu}  \, . \label{eq:frozenphi_final}
\eeq
In principle, then, in a matter-dominated universe, the field would eventually 
freeze to a value $\phi_m$, given by Eq.~(\ref{eq:frozenphi_final}). In 
practice, however, we find that the time at which this would start to happen, 
$t_\eq$ (defined in Eq.~(\ref{eq:stitchingeqsol})) is very large (much greater 
than the age of the Universe today). This is not a problem, though, as it
just means the field does not typically freeze during matter domination, 
instead slowly-rolling towards, but not reaching, $\phi_m$. Following this, 
once matter domination eventually gives way to dark energy domination, the field
dynamics will be determined by the matter-free equations of motion and the field
will once again tend to freeze at a value $\phi_s$ specified by 
Eq.~(\ref{eq:static_solution}). This process will, however, only complete itself
in the future of our present Universe when 
$\Omega_m\ll\Omega_{\text{DE}}\simeq 1$. At present, the field is envisaged to 
be in a state of slowly rolling towards $\phi_s$, due to the friction of the GB 
coupling and smallness of $\dot{\phi}$.

\subsection{Dark Energy Today}\label{sec:field_today}

Having confirmed that the behaviour of the field is sensible in the 
matter-dominated epoch, we can proceed to estimate the value it takes on today. 
We could have taken an approach where we would model the post-matter-domination 
evolution with some differential equation and evolve the Universe in time until 
the present day, essentially continuing to use the time coordinate $t$ to 
parametrise our position in the Universe's history. But instead, we use the 
known values of the present day dark energy and matter density parameters to 
this end. In this picture, the past state of perfect matter domination assumed 
in the calculations of section~\ref{sec:model_DE} is when $\Omega_m = 1$ and
$\Omega_\Lambda = 0$, the future dark energy domination which is eventually
reached by this model as matter dilutes and it tends to the static solution in 
Eq.~\eqref{eq:static_solution} corresponds to $\Omega_\Lambda=1$ and $\Omega_m=0$, 
while the present day values $\Omega_\Lambda\approx 0.7$ and 
$\Omega_m\approx 0.3$ indicate exactly where between these two limits we must 
presently lie. We are thus treating the density parameters as effective `time' 
coordinates to specify our point in the Universe's history. 

To formalise this, first note that the effective equation of state of the Universe $w = p/\rho$ and the derivative of the Hubble Parameter are related via the second Friedman equation in Eq.~(\ref{eq:FriedmanGB2}) such that
\beq
\mpl^2\dot{H} = - \frac{1}{2} (\rho + p) = - \frac{1}{2} (1 + w) \rho = 
- \frac{3}{2} (1 + w) H^2\mpl^2 \, .
\eeq
This can be used to rewrite the Klein-Gordon equation for the Gauss-Bonnet 
coupled field given in Eq.~\eqref{eq:GBKGE} as

\beq
\ddot{\phi} + 3 H \dot{\phi} + V_{,\phi} - 6 H^4  (1 + 3 w) G_{,\phi} \mpl^2 = 0
 \, ,
\eeq
which, under the slow-roll approximation $\ddot{\phi}\simeq 0$, is approximated 
by
\beq
\dot{\phi} \approx 2 H^3 (1+3w) G_{,\phi} \mpl^2  - \frac{V_{,\phi}}{3 H} \, . 
\label{eq:SRAsolw}
\eeq
Substituting this into the Friedman equation in the form
\beq
3 H^2 \mpl^2 \Omega_\Lambda = \frac{1}{2} \dot{\phi}^2 + V + 
12 \mpl^2 H^3 G_{,\phi} \dot{\phi} \, ,
\eeq
where $\Omega_\Lambda$ is the dark energy fraction, we obtain the approximate 
constraint equation

\beq
V + \frac{V_{,\phi}^2}{18 H^2} + \spar{3 \Omega_\Lambda + 
\frac{2}{3}(7+3w)V_{,\phi}G_{,\phi}}\mpl^2H^2+2(1+3w)(13+3w)(\mpl^2G_{,\phi})^2 
H^6 = 0 \, , \label{eq:phiconstr}
\eeq
which can be rewritten in terms of the explicit potentials of our model  
(assuming $\n \phi \ll 0$) in the form

\begin{align}\label{eq:phi_DE}
V_0 e^{2\n\phi_\de/\mpl} & + \frac{2\n^2V_0^2}{9H^2 \mpl^2} e^{4\n\phi_\de/\mpl} + 
2 \m^2 G_0^2 \mpl^2 H^6 (1+3w)(13+3w) e^{-2\m\phi_\de/\mpl} \nonumber \\
& +  \frac{4}{3} \m \n G_0 V_0 H^2 (7+3w) e^{(2\n-\m)\phi_\de/\mpl}  - 
3 H^2 \mpl^2 \Omega_\Lambda = 0 \, .
\end{align}
Under an appropriate substitution, this can be reduced to a polynomial equation 
for more straightforward algebraic or numerical analysis (though as the numbers
involved span many orders of magnitude, high-precision arithmetic should be used
in the case of numerical evaluation). Regardless of the preferred method, 
though, this constraint can be solved for $\phi_\de$ to identify the field value
necessary to achieve a specific equation of state $w$, dark energy fraction 
$\Omega_\Lambda$ and expansion rate $H$ for a given model, specified by $\n$ and 
$\m$. Assuming that following matter-domination, the Universe contains only 
matter and the dark energy field, we have from observations that 
$w = w_\Lambda \Omega_\Lambda \approx -0.7$, and $H_0 \approx 10^{-60} \mpl$. We 
also note that one can check that with the specifications $w = -1$, 
$\Omega_\Lambda = 1$ and $3 H^2 \mpl^2 = V$, representing perfect dark energy 
domination, solutions of Eq.~(\ref{eq:phiconstr}) yield $\phi = \phi_s$ as in 
Eq.~\eqref{eq:static_solution}, as expected, and in this limit 
Eq.~(\ref{eq:SRAsolw}) unsurprisingly reduces to $\dot{\phi} = 0$. \\

Interestingly, we typically find for most parameters that the $\phi_m$ value 
calculated in the previous section is larger in magnitude (more negative) than 
the $\phi_\de$ value obtained in the above procedure. This is consistent with 
the idea that $\phi_m$ is not in practice reached, and the maximum displacement 
of the field during matter domination is hence smaller in magnitude than 
$\phi_m$, as well as $\phi_\de$, and the field simply continues to roll and 
eventually reach $\phi_\de$ today. Alternatively, this could represent that the 
field \emph{does} overshoot $\phi_\de$ (though still not $\phi_m$) but then 
turns around due to the impeding effects of the GB coupling. This latter 
explanation is preferred, as earlier-time solutions of Eq.~(\ref{eq:phi_DE}) 
with $\Omega_\Lambda < 0.7$ are typically found to be larger in magnitude than 
the $\phi_\de$ value today at $\Omega_\Lambda = 0.7$. Furthermore, $\phi_s$, which
is achieved at later times when $\Omega_\Lambda \rightarrow 1$, is smaller in 
magnitude (less negative) than $\phi_\de$ today in the particular cases we have 
investigated in depth. These observations seem to suggest that the field is 
rolling `backwards' during the transition between matter and dark energy 
domination,but this may not be true for all models; we have not excluded the 
possibility that some parameters may lead to the change in direction only 
occurring after $\phi_\de$, or not at all.


\section{Constraints} \label{sec:constraints}

Having established the nature of inflation, reheating and dark energy in the 
framework of our model, we now proceed to use our results to constrain the 
parameter space to realistic values.

\subsection{Constraints from Inflation} \label{sec:constraints_inflation}
 
One can weakly constrain the parameter $\n$ from CMB constraints 
\cite{Planck2015Inflation} via the prediction of the spectral index in 
Eq.~(\ref{eq:i_ns}), which imposes roughly $\n\gtrsim 0.1$ when $N_*\approx 60$.
The spectral index obtained is $n_s = 0.9678$, which is in excellent agreement 
with observations, but this is not surprising as our model is explicitly designed to have an 
inflationary plateau. 
The tensor to scalar ratio calculated from Eq.~(\ref{eq:i_ttsr}) is, meanwhile, 
compatible with current constraints \cite{Ade:2015tva} for all values of $\n$  
(for $N_* \approx 60$ the maximum value of $r$ is around $0.03$ for $\n$ about 
$0.06$), but is mentioned here for completeness. The main constraint arising 
from inflation is the normalisation of the primordial power spectrum's 
amplitude. Using Eq.~(\ref{i_ppsa}) and imposing 
$\mathcal{P}_\mathcal{R} \approx 2.2 \times 10^{-9}$ \cite{Planck2015Inflation} 
gives 

\begin{equation}
V_0 \approx \frac{6.5 \times 10^{-8}}{\n^2 N_*^2}  \mpl^4  
\approx \frac{1.8 \times 10^{-11}}{\n^2}  \mpl^4 \, , \quad (N_* \approx 60) 
\, . \label{eq:V0constraint}
\end{equation}

The prefactor of the potential is hence set by inflation. To bring the potential
energy down to the dark energy scale at late times, a large suppression of many 
orders of magnitude is hence necessary. This implies we will need a rather large
value of $\n$ (much larger than the weak constraint the spectral index gives 
us). Just as a first estimate at this stage, noting that for $\n \phi \ll 0$ 
(deeply post-inflation) the form of the potential in 
Eq.~(\ref{eq:potentialtanh}) is approximated by $V_0\exp\rpar{2\n\phi/\mpl}$, 
we can see that for a maximum field displacement of $O(\mpl)$, we will need 
$2\n \approx \ln (V_0 / \Lambda) \approx O(100)$ to facilitate this.
We also note here that around this expected value of $\n \simeq 100$, 
Eq.~(\ref{eq:V0constraint}) implies $V_0 \simeq 10^{-15}\mpl ^4$, or 
$V_0^{1/4} \simeq 10^{14}\mathrm{GeV}$, close to the energy scale of 
Grand Unification.

\subsection{Constraints from Reheating and Dark Energy} 
\label{sec:constraints_reheating_DE} 

\begin{figure}
	\includegraphics[width=\textwidth]{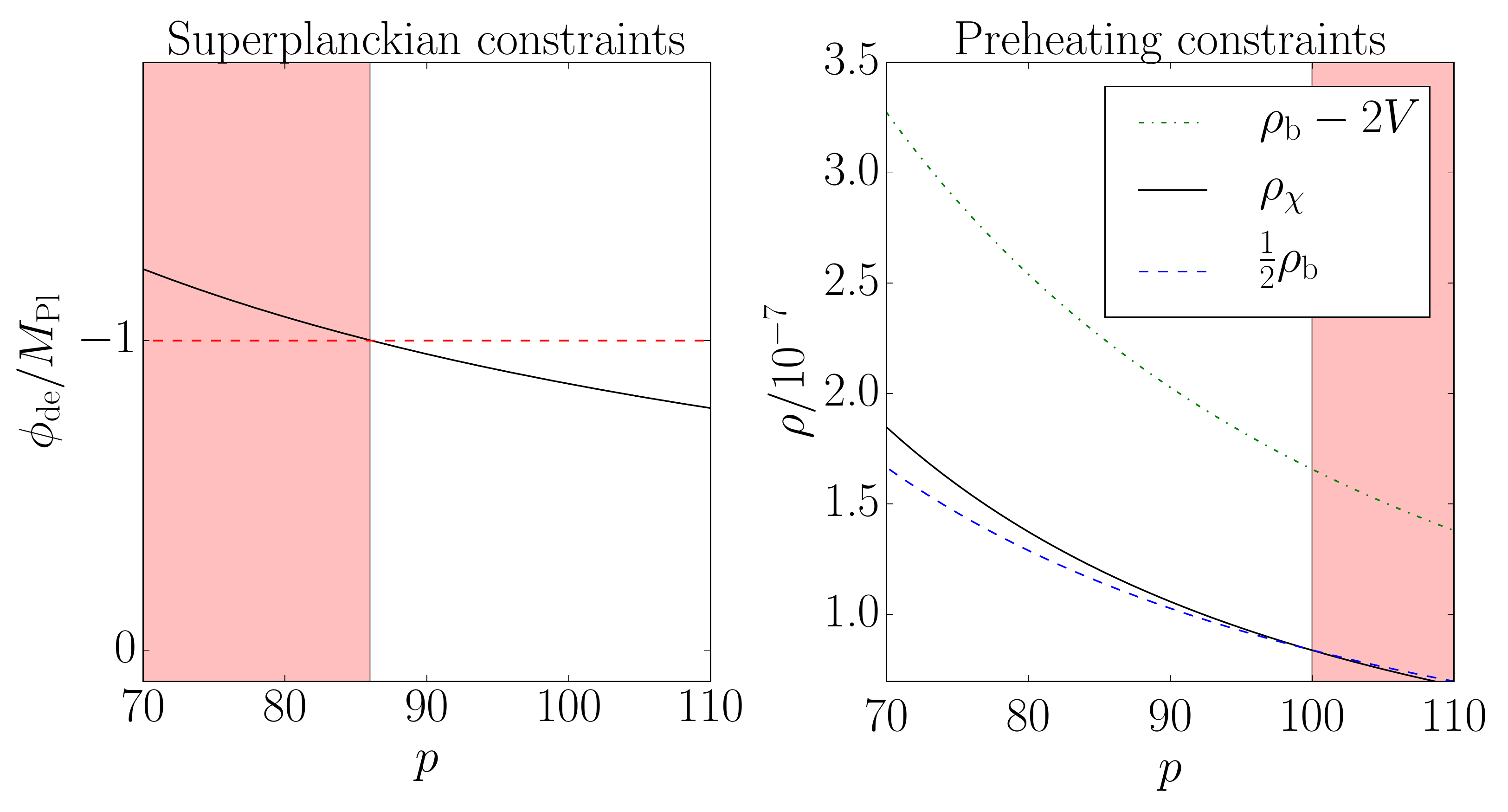}
	\caption{The left window shows the value of $\phi_\de$ computed via the methods of section~\ref{sec:field_today} for a range of $\n$ values when $\m = 4\n$, $G_0 \mpl^2 =1$ and $g = 0.8$. The shaded region on the left represents the parameter space where $\phi_\de$ is super-Planckian, and hence sensitive to details of unknown UV physics, imposing a bound of $\n > 86$ in this model. The right window shows the energy densities involved in the instant preheating conditions of Eq.~(\ref{eq:reheatingconditions}). The black solid line represents $\rho_\chi$, the density of radiation produced at instant preheating, while the blue dashed line and green dot-dashed line respectively represent the lower and upper bounds that $\rho_\chi$ must lie between. The shaded region on the right encloses the $\n$ values for which these inequalities are violated and hence imposes a bound of $\n < 100$ on this model.}
	\label{fig:Constraint}
\end{figure}

We numerically integrate the background equations of motion during inflation for
a range of models (specified by their  $\n$, $\m$, g and $G_0$ values). Using 
these results we compute the energy density resulting from instant preheating, 
the behaviour of the field in matter domination, the field value today, and the 
far-future field value $\phi_s$. We then impose the following constraints:

\begin{itemize}
\item Instant preheating should satisfy the conditions in 
Eq.~(\ref{eq:reheatingconditions}) for a sensible choice of the perturbative 
coupling to matter $g$ (i.e. $g \leq 1$).
\item The field value today, $\phi_\de$, should be subplanckian 
($|\phi_\de|<\mpl$) such that unknown UV physics do not strongly influence our 
results.
\end{itemize}
Examples of these constraints as a function of the parameter $\n$ for the case 
$\m = 4\n$, $G_0 \mpl^2 =1$ and $g = 0.8$ are shown in 
Figure~\ref{fig:Constraint}. The former shows that 
we must have $p > 86$ to avoid super-Planckian field values today, while the latter shows that instant preheating may not proceed according to the 
requirements in Eq.~(\ref{eq:reheatingconditions})  unless $\n < 100$. The 
resulting allowed parameter space of $86 < \n < 100$ for this case, alongside 
many other models with different $\m$, $G_0$ and $g$ values, is tabulated in 
Table~\ref{table:constraints}.

\newcolumntype{C}[1]{>{\centering\let\newline\\\arraybackslash\hspace{0pt}}m{#1}}

\begin{table}[]
	\centering
	
	\begin{tabular}{|C{2cm}|C{2cm}|C{2cm}|C{5cm}|}
		\hline
		$G_0 \mpl^2$         & $\m/\n$            & $g$ & $\n$ limits\\ \hline
		
		\multirow{6}{*}{1}   & \multirow{3}{*}{4} & 0.8 & 86 $ < \n < $ 100   \\ \cline{3-4} 
		&                    & 0.9 & 86 $ < \n < $ 238   \\ \cline{3-4} 
		&                    & 1   & 86 $ < \n < $ 507     \\ \cline{2-4} 
		& \multirow{3}{*}{8} & 0.8 & 51 $ < \n < $ 100   \\ \cline{3-4} 
		&                    & 0.9 & 51 $ < \n < $ 207   \\ \cline{3-4} 
		&                    & 1.0 & 51 $ < \n < $ 370   \\ \hline
		
		\multirow{6}{*}{100} & \multirow{3}{*}{4} & 0.8 & 85 $ < \n < $ 100    \\ \cline{3-4} 
		&                    & 0.9 & 85 $ < \n < $ 238    \\ \cline{3-4} 
		&                    & 1.0 & 85 $ < \n < $ 507   \\ \cline{2-4} 
		& \multirow{3}{*}{8} & 0.8 & 51 $ < \n < $ 72   \\ \cline{3-4} 
		&                    & 0.9 & 51 $ < \n < $ 155   \\ \cline{3-4} 
		&                    & 1.0 & 51 $ < \n < $ 258   \\ \hline
		
	\end{tabular}
	\caption{Table showing limits on $\n$ in the theory for various cases of the size of $G_0$, $\m$ and $g$, due to constraints coming from sub-Planckian field displacements and instant preheating's efficacy. In each case, the lower bound on $\n$ occurs as, below this threshold, $\phi_\de$ would have to undergo a super-Planckian displacement to serve as dark energy today. Similarly, the upper limits on $\n$ arise as, above these limits, the inequality in Eq.~(\ref{eq:reheatingconditions}) is violated. } 
	\label{table:constraints}
\end{table}

\subsection{Constraints from Tests of Modified Gravity} \label{sec:constraints_tests}

As our theory is a modification of General Relativity, it has to pass local 
gravity tests. Some work has been done on Gauss-Bonnet mediated dark energy 
theories in this context \cite{Sotiriou:2006pq,Amendola:2005cr,Amendola:2007ni}, with the 
strongest constraint found to be
\beq
\mpl |G_{,\phi}| \lesssim1.6 \times 10^{20}  \ \si{\square\metre} \approx 
1.5 \times 10^{88} \mpl^{-2} \, , \label{eq:cassiniconstraint}
\eeq
coming from constraints on the PPN parameter $\gamma$ based on the Cassini 
spacecraft's measurements of time delay of electromagnetic signals in a 
gravitational field. However, the analysis was made based on an assumption that 
derivatives of the scalar potential $V$ are all $O(V)$, which is strictly not 
applicable to our model in which $V_{,\phi} / V$ is $O(\n)$, and $\n$ is 
necessarily large. While a more comprehensive analysis of the case 
$V_\phi/V\gg1$ is necessary to work out the predictions for local gravity tests 
in our model, we will still briefly compare our results to this constraint to gain an approximate idea of what degree of constraint violation we may expect.

As in our model $\mpl |G_{,\phi}|  = \m G_0 e^{-\m \phi/\mpl}$, we can see that local
constraints of this form do not strongly constrain $G_0$, which only appears 
linearly in the constrained expression, but instead place a more stringent bound
on $\m$ which appears in a highly non-linear way in the expression above. 
Therefore, a relatively small increase in $\m$ is more likely to violate the 
constraint than an increase in $G_0$ by an order of magnitude. The constraint in
Eq.~\eqref{eq:cassiniconstraint} can be written as

\beq
\m \lesssim -\frac{\mpl}{\phi_\de}\,
W\rpar{\frac{1.5 \times 10^{88}}{G_0 \mpl^2}} \, . \label{eq:constrquant}
\eeq
Taking a Planckian field displacement $|\phi_\de| = O(\mpl)$ (corresponding to the lower-bound on $\n$ in parameter space) and $G_0 \mpl^2 = 1$ for argument's sake, we find $\m \lesssim 200$. However, in the best case scenario using the results from Table \ref{table:constraints}, $\m \gtrsim 340$. Similarly for models with larger $\n$, and hence $\phi_\de$ closer to zero, regardless of if we choose $\m = 4 \n$ or $\m = 8 \n$, we find that the constraint (\ref{eq:constrquant}) is up to a factor of around 2 smaller than the actual $\m$ value needed. This implies that there may be some basic difficulty in finding models which obey local gravity constraints while also satisfying the necessary criteria for dark energy and instant preheating, though the $\m$ value needed is at least comparable to (i.e. less than an order of magnitude larger than) the limit, such that if the constraints are moderately weakened when allowing for $V' \gg V$, there is some chance of the model passing local tests without further modification.

As our model is largely prototypical, particularly when choosing the shape of the potential, it is also feasible that modifications such as the addition of a screening mechanism to the model may also help alleviate any tension with local tests. We do not perform a thorough assessment of this now, however, because once again we emphasise that this constraint is based on assumptions that do not strictly apply to our theory. Analysis of this matter using more fit-for-purpose constraints is left to a future work.

\section{Discussion and Conclusions} \label{sec:conc}

We have studied in detail a model of quintessential inflation where the 
inflaton field couples to the Gauss-Bonnet (GB) term. By design, the GB coupling
is negligible at early times so inflation proceeds under standard slow-roll. 
Hence, we have considered a scalar potential, which features an inflationary 
plateau, as favoured by the latest CMB observations. Indeed, the scalar spectral
index found, \mbox{$n_s=0.9678$}, is close to the sweet spot of Planck 
observations, and we find a tensor amplitude considerably below the current upper bounds. 

As usual in quintessential inflation, the inflationary model is 
non-oscillatory, so that the inflaton field does not decay after the end of 
inflation, because it must survive until today to become quintessence and 
thereby explain dark energy observations without the need for an extremely 
fine-tuned cosmological constant. As the Universe must be reheated by means 
other than inflaton decay, we have employed the instant preheating mechanism,
in which the field is coupled to some other degree of freedom, such that as 
the field is rapidly rolling down the so-called quintessential tail of its
runaway potential, it induces massive particle production, which transforms
much of the kinetic energy density of the inflaton to the newly created 
radiation bath of the hot big bang. Soon afterwards, the inflaton field freezes 
at some value with small residual energy density, which becomes important at 
present, playing the role of dark energy. It is important to note that, while 
the field is rolling down the quintessential tail, it is oblivious to the form 
of the scalar potential, so our choice of model only determines the value of 
the residual potential density, when the field freezes. 

Because of the huge difference between the energy density scale of inflation 
and the current energy density (which is over a hundred orders of magnitude)
the inflaton field typically rolls over super-Planckian distances in field space,
in conventional quintessential inflation. However, this can result into a 
multitude of problems. Firstly, the flatness of the quintessential tail may be 
lifted by radiative corrections. Also, because the associated mass is so small,
the quintessence field may give rise to a so-called 5th force problem, 
which can lead to violation of the equivalence principle. To avoid these 
problems, it is desirable to keep the field variation sub-Planckian. In this 
case, however, to bridge the huge difference between the inflation and dark 
energy density scales, the quintessential tail must be steep. But, if the 
quintessential tail is too steep, when the field becomes important today, it 
unfreezes and rolls down the steep potential not leading to accelerated 
expansion at all. 

One way to overcome this problem is to make sure the scalar field remains frozen
today even though the quintessential tail is steep. To this end, in this paper 
we have considered coupling the field with the Gauss-Bonnet term, because such coupling 
impedes the variation of the field even if the potential is steep\footnote{%
For a different solution to this problem see Ref.~\cite{Dimopoulos:2017zvq}.}. Thus, in our
model the GB coupling becomes important at late times and makes sure the field
freezes with sub-Planckian displacement, such that it becomes the dark energy 
today without the aforementioned problems.

Quintessence is motivated only if the required tuning of the model parameters
is less than the extreme fine-tuning of the cosmological constant in 
$\Lambda$CDM. Quintessential inflation resolves one of the tuning problems of 
quintessence, that of its initial conditions, which are determined 
by the inflationary attractor. This means that the coincidence requirement
(that is that the dark energy must be such that it dominates at present) 
is satisfied only by virtue of the choice of the model parameters (and not by 
initial conditions). In our model, we have four model parameters, which account 
for the requirements of both inflation and quintessence. For the GB coupling, 
shown in Eq.~\eqref{GB}, we assume a simple exponential dependence on the 
inflaton, which ensures the GB coupling becomes important only at late times. 
The scale of the coupling is \mbox{$G_0\geq\mpl^{-2}$}, which agrees with our 
effort to stay sub-Planckian. For our scalar potential, shown in 
Eq.~(\ref{eq:potentialtanh}), the density scale is set by the COBE constraint 
to be \mbox{$V_0^{1/4}\sim 10^{-14}\,$GeV}, close to the scale of grand 
unification. In the exponent of the GB coupling and the argument of the $\tanh$
in the scalar potential, the inflaton field is suppressed by a large mass scale
$\mpl/\m$ and $\mpl/\n$ respectively. We considered \mbox{$\m\sim\n$} and found
that \mbox{$50\lesssim\n\lesssim 500$} (cf. Table.~\ref{table:constraints}), 
which means that, in both the GB coupling and the scalar potential, the inflaton
field is suppressed by the scale of grand unification $\sim 10^{16}\,$GeV. Thus,
we see that our model parameters are nowhere near tuned to the level of the 
extreme fine-tuning of the cosmological constant in $\Lambda$CDM. 

In summary, we have studied quintessential inflation where a coupling between
the inflaton field and the Gauss-Bonnet (GB) term allows the scenario to work, 
avoiding a super-Planckian variation of the field, which may otherwise be 
problematic. We considered a scalar potential with an inflationary plateau 
(favoured by CMB observations) and an exponential quintessential tail. Since 
the form of the potential is largely unimportant after inflation, we believe 
that our results are indicative for many quintessential inflation models with a 
GB coupling. We found that the model is successful for natural values of the 
model parameters, both in generating inflationary observables and also in 
accounting for the observed dark energy.

\begin{acknowledgments}
CvdB and KD are supported (in part) by the Lancaster-Manchester-Sheffield 
Consortium for Fundamental Physics under STFC grant: ST/L000520/1.
CL is supported by a STFC studentship. CO is supported by the FST of Lancaster 
University. 
\end{acknowledgments}

\bibliography{GBQuintrefs_final}

\end{document}